\documentclass[twocolumn,showpacs,preprintnumbers,amsmath,amssymb]{revtex4-2}
\usepackage{graphicx}% Include figure files
\usepackage{dcolumn}% Align table columns on decimal point
\usepackage{bm}% bold math
\usepackage{epstopdf}
\newcommand\ed{\tilde{\delta}}

\begin{document}
\title{Getting the tiger by the tail: Probing the turnaround radius of structures with outer
halo density profiles}
\author{$\text{Kyriakos Grammatikos}^{1,2}$ and $\text{Vasiliki Pavlidou}^2$} 

\affiliation{$ ^{1}\text{B}$oston University, Department of Physics, Boston, MA}

\affiliation{$ ^{2}\text{D}$epartment of Physics and Institute for Theoretical and
  Computational Physics, University of Crete, Greece}

\date{\today}
\begin{abstract}
In this paper we develop a new semianalytical approach to quantifying the density profile of outer dark matter halos, motivated by the remarkable universality those profiles, for a wide range of dynamical parameters of the inner halos. We show that our minimalist model is robust under significant variations of its input parameters around the currently known values and we identify the turnaround radius as the most important scale of the problem. Based on that observation, we argue that the turnaround radius accurately represents the transition to the external density profile, and we provide a universal test based on geometrical characteristics of the outer profile as a proposal for measuring the turnaround radius of structures.
\end{abstract}
\maketitle
\section{Introduction}
The advent of studies in the field of the formation of structures in our universe was a milestone in cosmology. Analyzing the way structures evolve in time and their state when they finally reach equilibrium can provide most unexpected insights into the state of the universe. 

One of many widely discussed problems in cosmology is the problem of the definition of a structure. In general, due to hardships posed by the non-linearity of Einstein's equations, the lack of a globally accurate description of the so-called "dark matter", but most importantly by the complexity of many-body problems in  general relativity, defining a structure has been non-trivial. Many notions regarding a structure are ill-defined. Even when the structure is in equilibrium, there is no concrete way of determining the boundaries of a structure (see \cite{press} and \cite{white} for a detailed discussion). Consequently,  measuring the total mass of a cosmological structure is also a poorly defined problem. In this context, other measurable quantities such as the density profile and the average radial velocity profile become important. 

 More specifically, the mass of a structure is a difficult quantity to define experimentally. In principle, it is easy to imagine that the mass of a structure corresponds to the mass of the maximum number of objects we can include in the structure that form a gravitationally bound system. As simple as this definition may seem, the lack of accurate information on the velocities and relative positions of the constituents of the structure demonstrates the difficulty of implementing the definition directly. Even theoretical calculations pose difficulties, due to the innate complexity structures exhibit as many-body systems. The result of those difficulties is that the prevailing physical scales used to describe structures are not always suitable for all kinds of phenomena. The necessity of accurate description of structure formation has therefore led to various proposals of scales which depend on the velocity or density profiles, which can substitute ambiguous concepts used up to present day.  

The first exact estimate for the density profile of a structure came from the pioneering work of Gunn \& Gott (1972), who utilized the spherical top hat model in order to approach in a simple way a highly non-linear problem. Since then and in the following few decades the field has known significant advancements. Earlier work in these steps was primarily focused on the density profiles of the innermost regions of structures as in \cite{nfw}. Progress in calculating the density profile of the outskirts of structures has been made only fairly recently. Furthermore, in the absence of observations for outer regions of halos, most effort was put in interpreting results for the innermost regions. However, these regions nowadays can be effectively probed by virtue of several methods (see \cite{halo}) , thus necessitating the theoretical interpretation of existing observations.

Numerical simulations, such as the ones presented by Diemer \& Kravtsov (2014), proved that the outskirts of simulated halos exhibit strong deviation from the commonly used density profiles of inner regions (NFW, Einasto), which manifests itself through a steep drop  in the power law locally describing the density profile over a narrow interval of radii. Additionally, their simulations revealed that the radial density profile tends to evolve almost self-similarly in time. Although it has been confirmed that there are several universal scalings related to external profiles, their physical significance and role are yet to be specified, since such studies along with semianalytical treatments cannot reveal whether this apparent universality is a well-established consequence of some physical aspect of the system. Characterization of halos has not been an easy task to tackle,  there is no rule of thumb for picking out the best criteria for detecting density peaks , and inevitably results ,especially concerning halo counting purposes, differ from author to author, depending on their particular choice of definition scale. For a discussion and comparison between several halo simulation algorithms see \cite{sim}.

Another issue is that the most important and actual physical changes in the behavior of the matter density profile of structure happen at lengths which are not described well by the standard scales used for the analysis of structure formation. Aanalysis of the spherical top hat model gives a concrete result for the radius of virialization of an idealized structure and predicts no other significant length scale for its density profile. More detailed cosmological simulations however contradict this result, proving that structures are far more complex, both inside and outside the virial radius. Important characteristics of the structures, such as peaks and second order critical points of the density profile, tend to appear around the virial radius, but also at greater distances, see \cite{krav} and \cite{ante}.Recent results \cite{krats} have shown that a reasonably robust choice of scale that distinguishes between regions where accretion is happening is the splashback
radius, and that this transition manifests itself as a steep drop in the logarithmic derivative of the density profile.

In the light of these considerations, several related questions arise: Does the category of spherical collapse models suffice for the description of radial density profiles of structures at large radii, where effects of non-sphericities are minimal? If that is true, could it possibly succumb to analytical methods and produce a prediction for the density profile of structures outside their virial radius that agrees with simulations? What physical mechanisms dictate the special characteristics of the external matter profiles? Is there any chance outer halo profiles are universal for all types of structures? Finally, at a more practical level, is there an easy and intuitive way for an experimenter to determine useful physical lengths of a structure, assuming that he is able to measure the outer radial density profile of a structure?

In this paper we calculate analytically the post virialization equilibrium density of a spherically evolving structure, which is assumed static. This analytical treatment is aimed at being reasonably accurate a considerable distance away from the center of collapse. We intend to draw conclusions from this treatment about the universality of the external matter profiles and determine explicitly on which parameters of the collapse they depend on. Also, we show that the scale that signifies the transition to the outer halo is no other but the well known turnaround radius of the overdensity. Finally, we propose a fitting function in the range of distances we are interested in, along with the best fit parameters of our approximate approach.

The paper is organized in three sections. In Section 2 we set up the idealized problem using arguments related to spherical symmetry. In Section 3 we derive analytically the halo density profile functional form. In Section 4 we present our results and discuss their implications, by comparing them to results of simulations. 

\section{Setup}\label{setup}
We consider at some early initial time, a background of a homogeneous and isotropic $\Omega_{m}+\Omega_{\Lambda}=1$ universe and at some point in space a spherical homogeneous structure of initial overdensity $\delta_{p}=\frac{\rho_{p}-\langle\rho\rangle}{\langle\rho\rangle}\ll 1$. In order to construct a spherical evolution model able to produce a radial density profile which evolves in time as well, we divide space into shells. Each shell is assigned its own average overdensity and subsequently evolves as a separate Friedmannian universe, affected only by the mass it encloses. This procedure is expected to produce a radial density profile of the structure the initial perturbation induces around it, at each future universal cosmological time, or equivalently at each value of the scale factor $a$ of the initially unperturbed universe. In the following, we will not consider shell crossing, as shell crossing will be of limited spacial extent. Our solution therefore is applicable only to radii outside the relaxed structure as well as the region of active accretion and mass redistribution. 

First we would like to assign an average overdensity to each shell. We define the average density at the initial time considered as the total mass inside the shell over the volume of the shell at radius r, and its average true overdensity denoted by $\langle\delta\rangle(r,a_{i})$ and defined by an analogous relation $\langle\delta\rangle(r,a_{i})=\frac{\langle\rho\rangle(r,a_{i})-\langle\rho\rangle}{\langle\rho\rangle}$  is found to be:

\begin{equation}
\langle\delta\rangle(r,a_{i})=\delta_{p}\Big(\frac{r_{p}}{r}\Big)^{3},
\end{equation}
where $r_{p}$ for the initial radial extent of the perturbation. To derive this relation we have taken into account that the rest of the universe outside the perturbation has the same density everywhere, $\langle \rho\rangle $, inside the overdensity matter density is $\rho_{p}$ and we had the following simple relations in mind, which can be derived if one keeps track of the total mass contents of the shell:

 $$\rho_{p}\frac{4}{3}\pi{r_{p}^{3}}=m_{p}$$
$$\langle\rho\rangle(r,a_{i})\frac{4}{3}\pi r^3= m $$
$$\langle\rho\rangle\frac{4}{3}\pi (r^3-r_{p}^3)=m-m_{p},$$

We would like to recast equation (1) in terms of the mass inside the shell $m$ and the mass of the perturbation $m_{p}$ only. This will be helpful since the mass inside each shell turns out to be an integral of motion if we enforce our no-crossing assumption. Combining equation (1) and the three relations we find that:

\begin{equation}
\langle\delta\rangle(r,a_{i})=\delta_{p}\frac{m_{p}}{m}\frac{1}{1+(\frac{m-m_{p}}{m})\delta_{p}}
\end{equation}
For typical values of the overdensity of the initial seed $\delta_{p}=10^{-3}$ and noting that outside the perturbation $\frac{m-m_{p}}{m}<1$, by making an error of order $\delta_{p}$ we obtain to first order:
\begin{equation}
\langle\delta\rangle(r,a_{i})=\delta_{p}\frac{m_{p}}{m}
\end{equation}
The linearized overdensity field is defined as the solution to the linearized equations governing the phenomenon. Our intention is to establish a relation between the true and linear overdensity field, since calculating the linear one is much easier. Again,defining the average linear overdensity as before $\tilde{\langle\delta\rangle}(r,a)$ and using the fact that it is proportional to a growth factor $D(a)$ given by \cite{pfd} (D18) we can easily prove that, since linear and true overdensities are identical at early times :
\begin{equation}
\tilde{\langle\delta\rangle}(r,a)=\tilde{\langle\delta_{p}\rangle}(a)\times\frac{m_{p}(a)}{m}
\end{equation}
which means that to calculate the linear overdensity at any epoch we need only monitor the evolution of the overdensity itself. What remains to be done after that is to link the values of true and linear overdensities to find the true overdensity, which is the quantity of interest. We would like to note here that the mass inside the core of the structure $m_{p}$ is also an integral of motion ,at least for as long as radial infall has not commenced. In that particular case the epoch dependence drops out completely from the mass of the perturbation.

\section{Formalism}\label{formalism}
It is inevitable that inside the 
overdensity, the surrounding mass shells will eventually start to fall to the center of the structure. At the time shell crossing takes place, especially around the time that the perturbation formally collapses to a singularity, virialization will begin to occur inside the structure through 3-body interactions, forcing infalling particles to attain non-zero angular momentum. For our model we will assume that the virialization of the structure happens exactly at the time of the formal collapse of the structure into a singularity. We will make the approximation that the structure remains virialized at all times, in order for us to be able to consider average spherical symmetry inside the structure. Real structures of course evolve differently due to the evolution of their inner core, which is responsible for all kinds of accretion and non-sphericities appearing in more detailed analyses. This assumption of equilibrium therefore is a primary source of error in our treatment. 

Now, for the epoch of virialization, we would like to obtain the radial density profile of the halo of the structure. We expect that this profile will be the steady-state profile of the outer halo of any virialized structure in the universe. We define the epoch for which the radius of the perturbation first vanishes 
to be the collapse time, $a_{coll}$. In principle, we could obtain the radial density profile by exploiting the integrability of a $\Lambda LTB$ model. This would result in an expression for the density profile involving vacuum integrals. However we would like to pursue a semianalytical method for determining the radial density profile at the time of virialization.

Applying relation (4) at the time of collapse (or virialization equivalently) we find that:
\begin{equation}
\tilde{\langle\delta\rangle}(m,m_{p},a_{coll})=\tilde{\delta}_{c}(a_{coll})\frac{m_{p}}{m}
\end{equation}
where $\tilde{\delta}_{c}$ stands for the value of the linearly extrapolated overdensity at the time of collapse.

In order to be able to make a measurement at the epoch of virialization,we must extract the linearized overdensity as a function of the comoving observer distance $R(r,a_{coll})$. Using equation (A8) we solve approximately for the true overdensity(we suppress all dependences but the ones referring to a specific shell):

\begin{equation}
\langle\delta\rangle(m)=\Bigg(1-\frac{\langle\tilde{\delta}\rangle(m)}{\tilde{\delta}_{c}}\Bigg)^{-\tilde{\delta}_{c}}-1
\end{equation}
Now we are able to find the true density contrast assigned to each shell as a function of the mass $m$ inside the shell and also the average density inside:
\begin{equation}
\langle\delta\rangle(m)=\Big(1-\frac{m_{p}}{m}\Big)^{-\tilde{\delta}_{c}}-1
\end{equation}
\begin{equation}
\langle\rho\rangle(m)=\rho(a_{coll})\Big(1-\frac{m_{p}}{m}\Big)^{-\tilde{\delta}_{c}}
\end{equation}
where $\rho(a_{coll})$ is the density of the unperturbed universe at the epoch of virialization. Since the mass inside the shell of comoving radius $R$ is an integral of motion, by the definition of the average density we assigned to each shell, we can find $R(m)$, the radius of the shell as a function of the total mass it contains:
$$m=\langle\rho\rangle\frac{4}{3}\pi R^3(m)$$
\begin{equation}
R(m)=\Big(\frac{4}{3}\pi\rho(a_{coll})\Big)^{-\frac{1}{3}}m^{\frac{1}{3}}\Big(1-\frac{m_{p}}{m}\Big)^{\frac{\tilde{\delta}_{c}}{3}}
\end{equation} 
Finally, to replicate the density profile of the structure, it will suffice to find the density as a function of shell mass content m, which will replace equally well the distance from the origin as the varying parameter.

Utilizing the definition of the total mass content of a shell one can show that:
$$\rho(m)=\frac{1}{4\pi R^2(m)\frac{dR(m)}{dm}}$$
from which, using equation (9), we obtain the desired result:
\begin{equation}
\rho(m)=\rho(a_{coll})\Big(1-\frac{m_{p}}{m}\Big)^{1-\tilde{\delta}_{c}}\frac{1}{1+(\tilde{\delta}_{c}-1)\frac{m_{p}}{m}}
\end{equation}

Defining $X=\frac{m}{m_{p}}$ and the length $L=\Big(\frac{3m_{p}}{4\pi\rho(a_{coll})}\Big)^{\frac{1}{3}}$ we obtain the parametric equations for the density profile as functions of $X$:
\begin{equation}
\rho(X)=\rho(a_{coll})\Big(1-\frac{1}{X}\Big)^{1-\tilde{\delta}_{c}}\frac{1}{1+(\tilde{\delta}_{c}-1)\frac{1}{X}}
\end{equation}
\begin{equation}
R(X)=LX^{\frac{1}{3}}\Big(1-\frac{1}{X}\Big)^{\frac{\tilde{\delta}_{c}}{3}}
\end{equation} 

Relations (11) and (12) are the main results of our model.

\begin{figure}
%\resizebox{3.3in}{!}{

\includegraphics[width=0.48\textwidth]{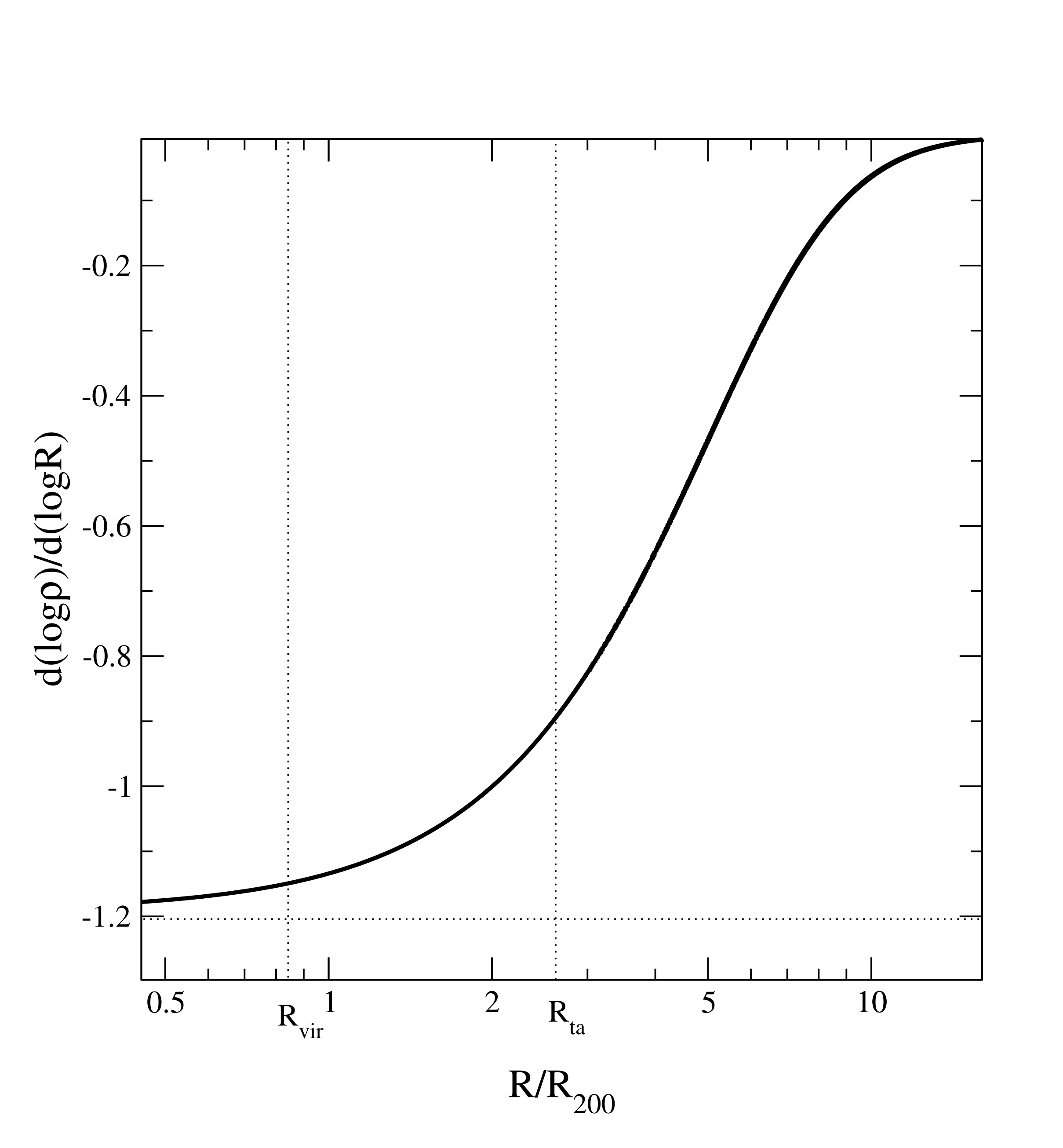}%}% Here is how to import EPS art
\caption{\label{fig:papergraph3} Plot of the logarithmic slope of the density profile, based on equations (11) and (12). By comparison with the simulations of \cite{krav} all structures, regardless of their specific parameters, have the same profile at about the turnaround radius at the epoch of virialization. After the second vertical dashed line, which represents the turnaround radius, we expect that all structures will exhibit the same radial profile as in our plot. We expect our model to be valid in that range, but not any closer than the turnaround radius. }
\end{figure}

\section{Results}\label{results}
The density profile we obtain from the above equations has subtle characteristics, which can only become apparent if we draw the logarithmic slope of the graph versus distance as in Fig. 1. As expected, our simple model, which does not take turnaround into account, presents us with a smoothed out but also non-trivial density structure of the surrounding universe.

First of all, we notice that the logarithmic slope of the density profile approaches asymptotically as $X\rightarrow1$  a well-defined limit:

$$\lim_{X\rightarrow1}\frac{d(\ln\rho)}{d(\ln R)}=3\frac{1-\ed_{c}}{\ed_{c}}\approx-1.2 ,$$
where the linear overdensity at the time of virialization is chosen to be $\ed_{c}=1.675$, which is the value theoretically predicted by the spherical collapse model, when the accepted values $\Omega_{\Lambda}=0.7, \Omega_{m}=0.3$ of the contents of our Universe are taken into account.

This limit depends only on the cosmology of the universe where the perturbation is embedded, and the specific properties of the model employed, and in this approximation independent from the mass and radius of the virialized structure.  Therefore we reach the conclusion that this limit exhibits universality in the sense that any spherical accretion instance, regardless of the initial conditions (mass and overdensity of the perturbation), will result in a power law density profile near the edge of the perturbation. The power law will be approximately $\rho\sim r^{-\frac{6}{5}}$ if we correctly assumed that the virialization epoch of the structure is the same as the epoch of formal collapse in the spherical top hat model. If we take shell crossing into account, we may expect that due to extensive mass accretion, this law will steepen near the structure. We stress out at this point that our model is not expected to give accurate results for the region over which the limit is taken, however we present this behavior as representative of the universality of the model.

Comparing the resulting logarithmic slope we calculated with simulations of Diemer \& Kravtsov \cite{krav} for the same physical quantity, we observe that all halos they present, independently from the parameters that characterize their accretion rate, or peak height, are described more or less by the same profile in their outer regions. In sufficiently long distances, and in particular when $R\simeq R_{ta}$, most halos, and in particular older ones, which are denser in the center, begin to look the same.

These observations lead us to the conclusion that the virial radius is not the only scale one needs for describing a structure. That is evident in [6] where the static region of a cluster-sized structure usually appears several
virial radii away. Gradients in the density field tend to appear even farther away. Although halos feature a wide range of evolution patterns, depending on initial conditions, the density profiles bear striking similarities as one moves to their outskirts. The turnaround radius of the structure seems to be strongly signifying the transition to the external parts of a halo, which fact renders it the dominant physical scale in that region. All this evidence is in support of an almost universal profile for outer halos and in the limit of perfect sphericity, our model should be valid.
Let us emphasize here that the turnaround radius is located far away from any active accretion events, as for the overwhelming majority of structures, any significantly accreting region exists approximately within 2 virial radii away from the center. The accuracy of the model is affected by the non-sphericities, which are most prominently present in newer structures, which have not yet virialized and restored maximal symmetry in their structure. Another significant source of error in our model is the assumption of equilibrium, or in other words, the fact that we assume that there is no shell crossing. This assumption effectively decouples the interior from the exterior, and thus we end up neglecting matter splashback towards the
exterior. Most notably, splashback creates primary and secondary dips in the density profile that make it globally steeper, especially for heavily accreting halos. The more active the interior is, the more our model loses predictive power of the density gradient, even at distances as far as $5R_{vir}$ for active halos. However, the niche of this
treatment is in that the prediction of the $\textit{inflection point}$ is robust, and as shown in simulations done by the previous references, almost universal. This opens up the possibility of using the inflection point appearing in the profile as an indicator of the transition to the halo exterior and a device of measurement of the turnaround radius.

\begin{figure}[!ht]
\includegraphics[width=0.48\textwidth]{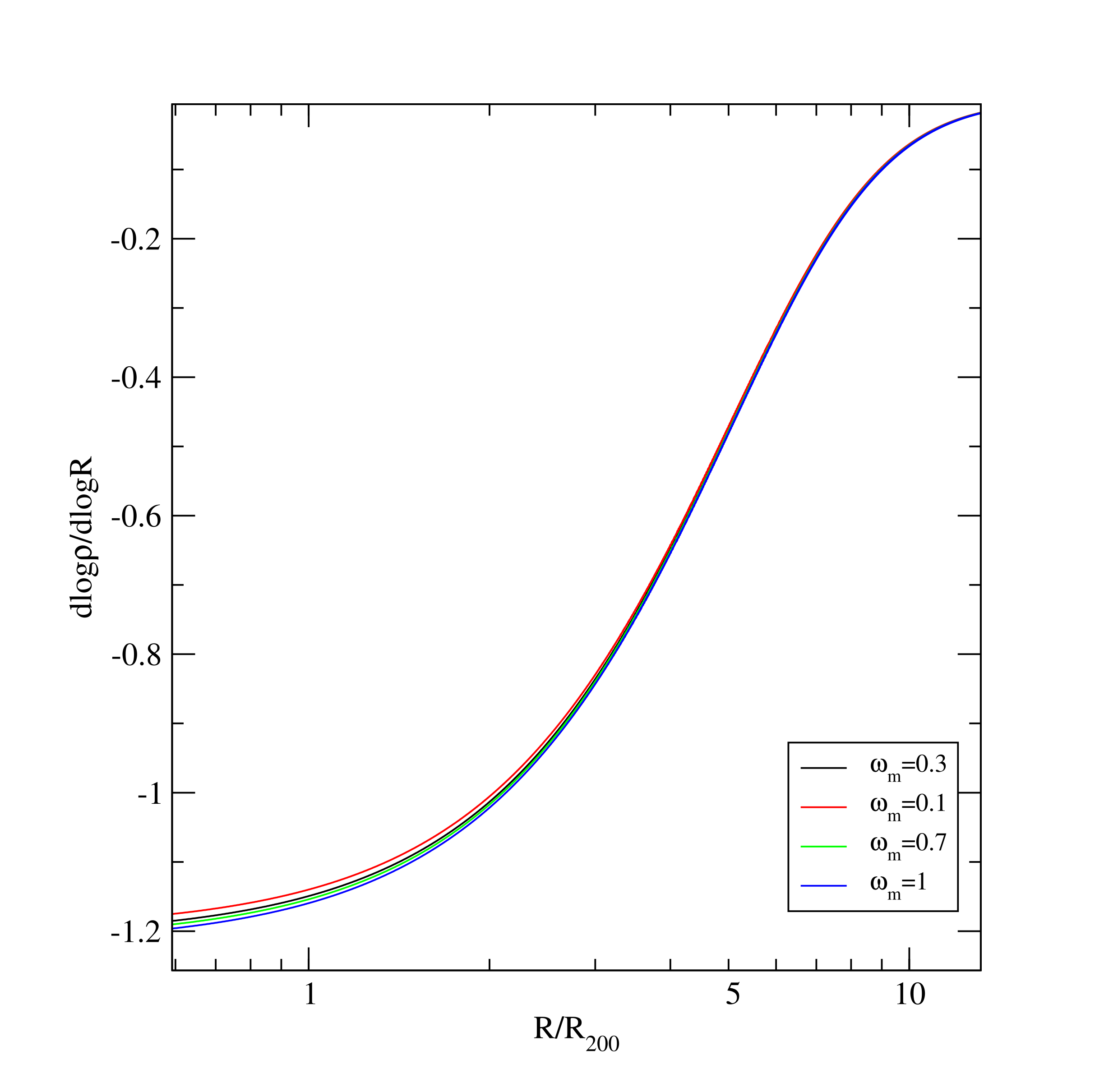}
\caption{\label{fig:cosmologydependence} Plot of our model for a number of different cosmologies. The model obviously exhibits insensitivity to the probed range of cosmologies. The outer halo profiles for each cosmology are very similar, especially closer to the edge of the structure. This implies that outer halo profiles seem to be a good choice of a universal way to obtain information for structures.}
\end{figure}

From the plots of Fig. 2 we can see that the variation of the outer profile with respect to the cosmology is obviously small. The outer halo density profile is insensitive to a very wide range of CDM contents. This fact solidifies the evidence we have for the universality of the profiles under question but it also prevents the use of information from outer halo profiles as an indicator of the ambient cosmology. Note however that in the context of the spherical collapse model, the size of the turnaround radius as a function of the structure's mass (see Eqn. A9) can be used as a discriminator of cosmology due to the dependence of the true overdensity at turnaround on it. %???????????????????? proxy identity

This universality allows us to introduce a setup that can help experimentalists calculate easily some characteristic length scales of the halo, using only our profile. In its range of applicability, our profile exhibits an inflection point. This inflection point can be used as an indicator from which we can calculate all physical scales of interest. The position of an inflection point can be inferred with decent accuracy from observational data due to its geometrical nature. Here we should stress out however that the position of the inflection point depends critically on the way the function is plotted. That means, that the inflection point depends on the metric function of the distance axis. In order to achieve agreement, we will find the inflection point for a logarithmically and a linearly plotted distance axis.

\begin{figure}
%\resizebox{3.3in}{!}{
\includegraphics[width=0.48\textwidth]{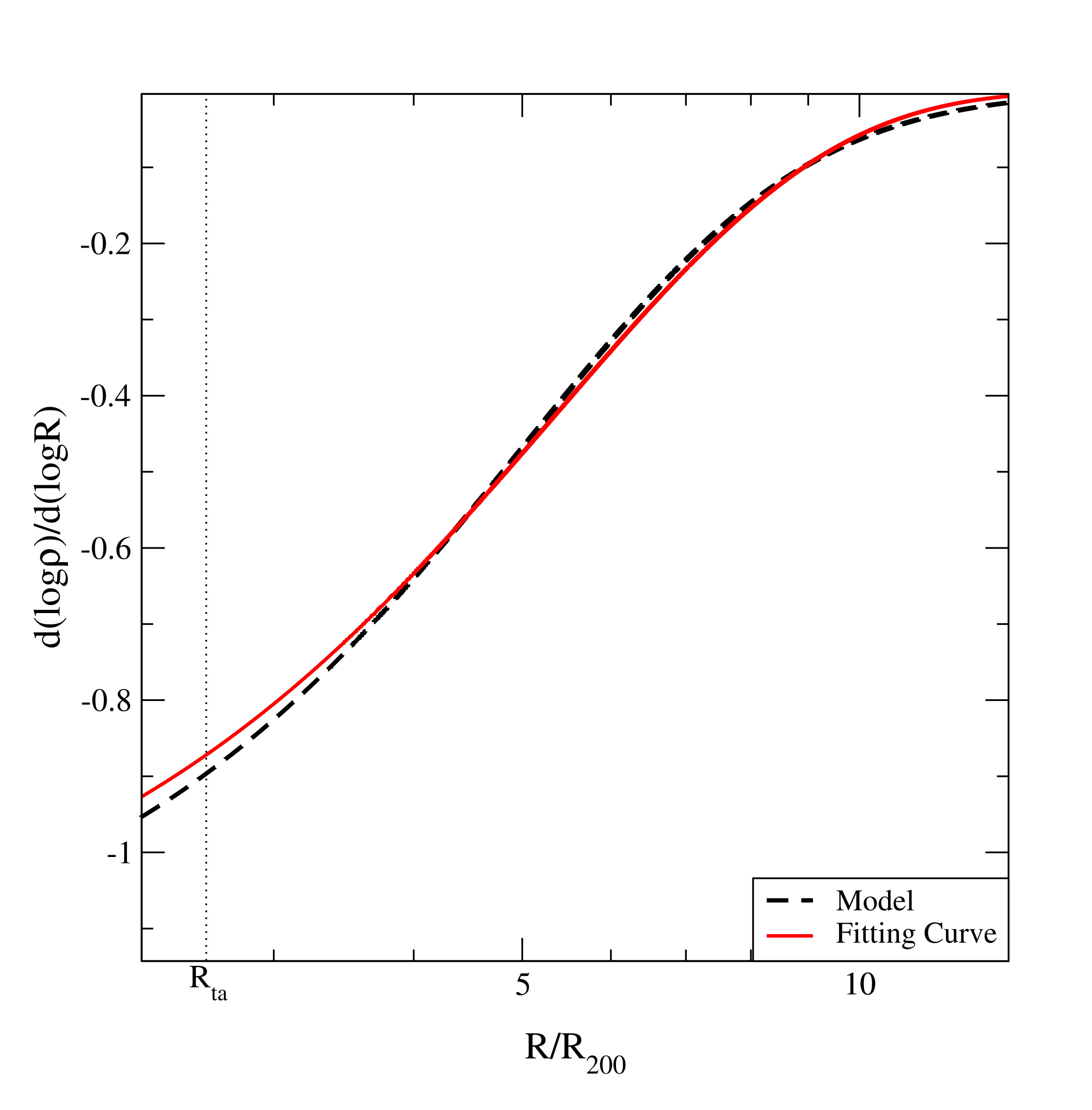}%}% Here is how to import EPS art
\caption{\label{fig:papergraph5} Comparison of the proposed fitting curve with the complete model. The best agreement between the two is achieved in the range $4-10R_{200}$. The length scale inserted in the error function is $R_{0}=1/\alpha= 1.13R_{200}$. }
\end{figure}

After calculating the second logarithmic derivative of Fig. 1 we find that the inflection point is to be found by solving the equation:

$$2X_{c}^2-(2\ed_{c}+1)X_{c}+(1-\ed_{c})=0$$

Solving the equation, we have, taking into account the value of the linear overdensity at the time of collapse just like before, and then plugging in the corresponding overdensities for the time of turnaround:
\begin{equation}
X_{c}=2.007, R_{ta}=0.506R_{c}  \text{ \hspace{0.5 cm}(logarithmic plot)}
\end{equation}
For reference, we also give the corresponding rule for the linear plot:
\begin{equation}
X_{c}=1.448, R_{ta}=0.739R_{c}  \text{\hspace{0.5 cm}(linear plot)}
\end{equation}

These calculations have been made for structures today, at redshift z=0. But what happens as we move back in time? As the redshift of a structure increases, its turnaround radius changes according to the spherical collapse model. An exact relation between true and extrapolated overdensities would also involve some kind of dependence on cosmological time. In our approximate relation, we consider that redshift dependence is weak and therefore negligible. Thus, the only change that takes place with redshift is the value of the turnaround radius each virialized structure possesses. This induces an easily calculable change to the relative position of the inflection point of the radial density profile (which remains constant in time) and the turnaround radius, which changes upon redshift variation.We present a graph that illustrates the variation of this relative position with different redshift values. For testing structures of different redshifts, the correct value has to be taken from the plot of Fig. 4 in order to estimate properly the turnaround radius.

The parametric equations derived for this profile seem to be difficult to manipulate for fitting purposes, and therefore we propose an analytical formula for modelling the halo. We found a two parameter model, presented in Fig. 3 which agrees at the 5\% level with the logarithmic slope for a wide range of distances. 
\begin{equation}
\frac{d\ln \rho}{d\ln R}=1.2((\text{erf}(\alpha R))^\beta-1)
\end{equation}

\begin{figure}[!ht]
\includegraphics[width=0.48\textwidth]{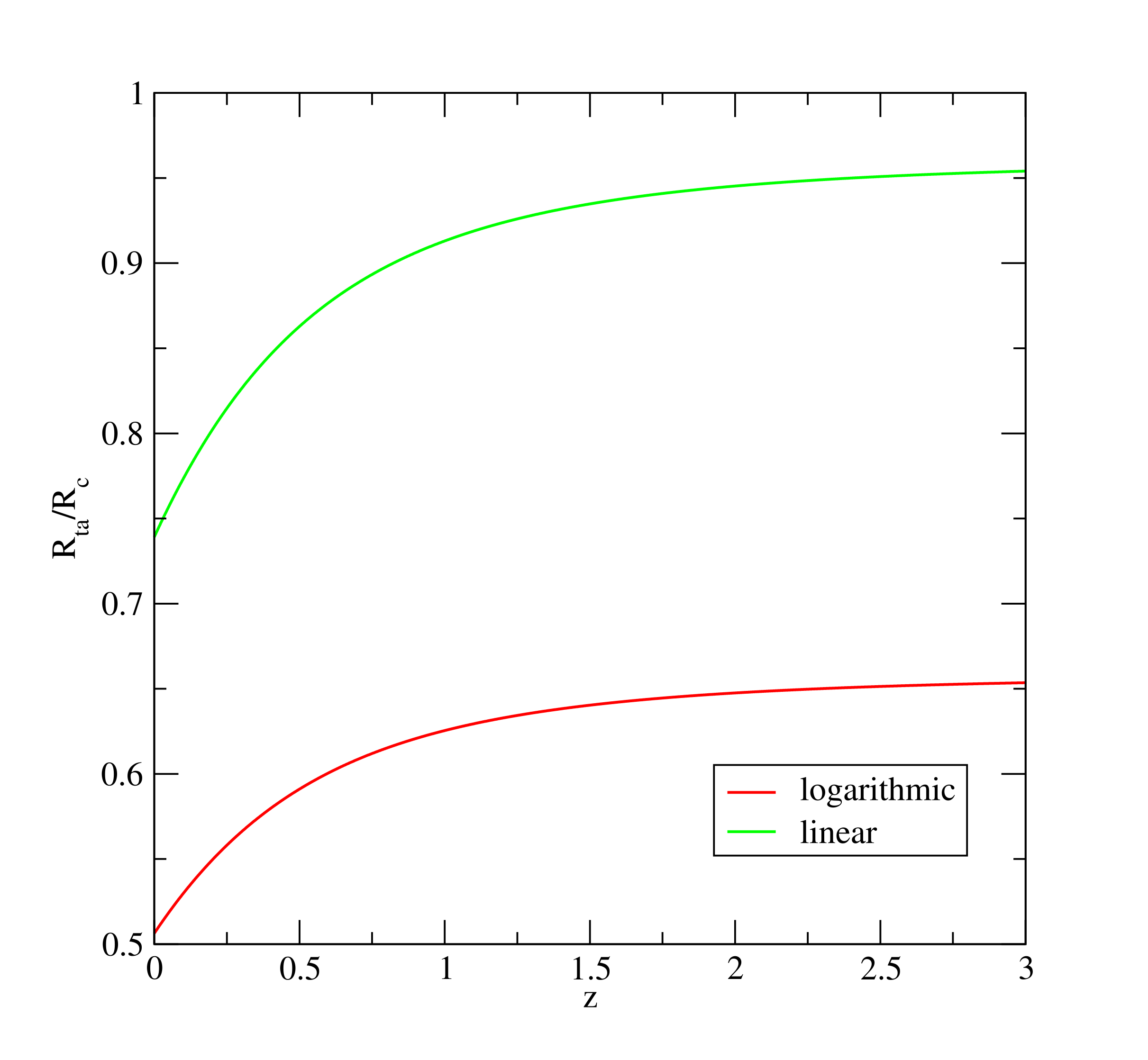}
\caption{The variation of the ratio of the turnaround to inflection point radius with redshift is presented. The red curve represents the relative position of the two radii for a logarithmic plot of the density profile, and the green curve for a linear plot. }
\end{figure}

with the best fitting coefficients being:

$$\alpha=0.151, \beta=1.513$$

\section{discussion}\label{discussion}
In this paper we have presented an analytical form for the radial density profile of the outskirts of a spherically symmetric structure at its time of collapse. We made use of the spherical collapse model and successful empirical relations between true and linear overdensity fields of such an object and our analysis revealed that in this approximation all structures appear similar in their outer
parts. This justifies the apparent universality seen in simulations of external halo matter profiles on theoretical grounds. Also, it emphasizes the very important role played in structure formation by the turnaround radius,
itself being the particular scale the outer halos remember about in their dark energy dominant phase of evolution. 

The physical mechanisms that lie in the roots of this phenomenon are illuminated. The turnaround radius of the core of the structure is the dominant length scale that signifies the transition to regions where mass infall becomes significantly weaker, and eventually the structure's halo constituents joins the Hubble 
ow.
We use the geometrical features of our profile as a way of determining significant radial scales of a structure and stress out the insensitivity of the outer halo profile to
the ambient cosmology. The close relation between the turnaround radius and those geometrical features provides us with a universal, redshift dependent probe based
on spotting the critical point of the outer halo density profile, that allows us to calculate the turnaround radius of a given structure from experimental data. Finally, we find an analytical empirical relation describing the slope of the density profile at 5\% accuracy in the desired range of radii.  

In general, the range of applicability of our analysis is limited by a number of factors: 1) Deviations from spherical symmetry which are more prominent before virialization and in smaller structures,
2)radial infall and therefore shell crossing occuring around the virial radius, which alters the way mass is distributed in the halo, and
3)pressure gradients which may occur if individual cosmological masses are subject to forces caused by any kind of interactions at zero temperature, such as the ones a barotropic fluid introduces. Our solution refers to a fluid which exerts no force to anything, but noone guarantees that a cosmological fluid behaves in this manner.

Despite the numerous limitations the class of $\Lambda$CDM models exhibits, our assumptions are not one of them. Discussion and justification of the approximations in the
appendices to follow aims to provide clarity and mathematical rigor. We will be striving for better understanding of the aforementioned issues and obtaining further qualitative and quantitative information about more
complicated models in future work.

\acknowledgments{We gratefully thank Dimitrios Tanoglidis, Theodore Tomaras and Konstantinos Tassis for enlightening discussions.}

\appendix
\section{the spherical collapse model}  
We would like to begin by stating the e
quation of motion we obtain for a single perturbation by applying the spherical collapse model:

\begin{equation}
\Big(\frac{da_{p}}{dt}\Big)^2=H_{0}^2\Omega_{m,0}\frac{\omega a_{p}^3-\kappa_{p}a_{p}+1+\delta_{0}}{a_{p}}
\end{equation}

,where $\delta_{0}$ represents the initial overdensity of the perturbation, and the value of $\kappa_{p}$ dictates the nature and intensity of the curvature of space inside the perturbation. This quantity and $\omega$ are related to initial conditions by the following relation:

\begin{equation}
\kappa_{p}=(1+\delta_{0})+\omega -\Big(\frac{H_{p,0}}{H_{0}}\Big)^2, \hspace{0.5cm} \omega=\frac{\Omega_{\Lambda ,0}}{\Omega_{m,0}}
\end{equation}

This equation can be slightly modified by setting $\tilde{a}_{p}=a_{p}(1+\delta_{0})^{-1/3}$ and $\tilde{\kappa}_{p}=\kappa_{p}(1+\delta_{0})^{-2/3}$, so that it can be handled more easily:

\begin{equation}
\Big(\frac{d\tilde{a}_{p}}{dt}\Big)^2=H_{0}^2\Omega_{m,0}\frac{\omega \tilde{a}_{p}^3-\tilde{\kappa}_{p}\tilde{a}_{p}+1}{\tilde{a}_{p}}
\end{equation}

Finally, the equation of expansion of the surrounding universe is given by:

\begin{equation}
\Big(\frac{da}{dt}\Big)^2=H_{0}^2\Omega_{m,0}\frac{\omega a^3 +1}{a}
\end{equation}

It will be useful to eliminate time measured by the clock of the comoving observer in favor of the cosmological time defined by the scale factor of the surrounding universe, to obtain the result:

\begin{equation}
\Big(\frac{da_{p}}{da}\Big)^2=\frac{a}{a_{p}}\frac{\omega a_{p}^3-\kappa_{p}a_{p}+1+\delta_{0}}{\omega a^3+1}
\end{equation}

If we want to create a radial density profile for the structure instead of studying the evolution of the perturbation boundary alone, we construct an equation of motion for every mass shell into which we divide the universe outside the perturbation. We do not consider what happens inside the perturbation. It is true that initially the perturbation will continue to be homogeneous during its expansion, as expected, but after infall commences, there is a possibility that shell crossing will occur and the density profile of the perturbation itself will change inside it. Here we will state an equation of motion for the scale factor of each shell, denoted by $G(r,a)$, which is valid before turnaround:

\begin{equation}
\Big(\frac{dG}{da}\Big)^2=\frac{a}{G}\frac{\omega G^3-\kappa\text{(r)}G+1+\delta_{i}\Big(\frac{r_{p}}{r}\Big)^3 }{\omega a^3 +1}
\end{equation}

where by use of the initial overdensity assigned to each shell, given by equation (1) we find that:

\begin{equation}
\kappa\text{(r)}=1+\omega-\Bigg(\frac{H_{0}\text{(r)}}{H_{0}}\Bigg)^2 +\delta_{i}\Big(\frac{r_{p}}{r}\Big)^3 
\end{equation}

The physical size of each shell at any given time is  \hspace{0.5 cm} $R(r,a)=rG(r,a)$ and the quantity $H_{0}(r)$ represents the initial expansion rate of each shell. The preceding equation can also be recast into the standard form of equation (A4) by similar scaling arguments as before.

This model is insufficient in describing spherical collapse when shell crossing occurs, since in that case the mass each shell encloses begins to gradually change, and thus the evolution is no longer similar to an independent Friedmannian Universe. Further exploration of models of spherical collapse reveal caustics in particle trajectories as in \cite{bertschinger}.

Furthermore, we would like to note that the essence of this semianalytical treatment is based on an empirically derived analytical relation, reliable on the 5 \% level in the desired range of overdensities. It allows us to translate analytical results from linear theory into statements about a non-linearly evolving structure. We used the relation stated in \cite{pfd} and also in earlier works (Sheth 1998).Explicitly, the following relation between linear and true average overdensities of any mass shell has been verified:

\begin{equation}
\ed_{a}=\ed_{c}\Big(1-(1+\delta_{a})\Big)^{-1/\ed_{c}}
\end{equation}  

In the treatment above, any characteristic length scale of the structure denoted by "x" is determined by the following equation:

\begin{equation}\label{eq:A9}
m_{p}=\rho(a_{coll})(1+\delta_{x})\frac{4}{3}\pi R_{x}^3 
\end{equation}

One can calculate the true overdensity of the structure at the time corresponding to the specific length scale by manipulating the corresponding spherical collapse model of equation (A5).

\section{an exact treatment using the LTB model}  

The LTB model is the most general spherically symmetric dynamical solution to the Einstein equations for a pressureless perfect 
uid in a universe with non-zero
cosmological constant, which is known to be completely integrable up to quadrature. It must be the case that more general spherically symmetric solutions exist, that allow for the formation of an event horizon. Elaborating on the LTB formalism offers a rough view on the
conditions required for our approximation to hold. We expect this model to be capable of describing structure formation/expansion in intermediate, non-violent stages of their evolution on grounds of its generality and the
fact that structures, although they are relatively small many-body systems, dark energy dominates their evolution enough[15], so that a pressureless perfect fluid description will capture the general picture of their evolution.

The Lemaitre-Tolman-Bondi metric ansatz is given by an expression of the form:
\begin{equation}
ds^2=-c^2dt^2+X^2(r,t)dr^2+A^2(r,t)(d\theta^2+\sin^2 \theta d\phi^2)
\end{equation}

As one can easily see, this ansatz for the metric does not allow for the formation of a black hole event horizon since the $g_{tt}$ component of the metric tensor is never
zero. As we will later see, assuming the universe only contains pressureless dust trivializes the radial evolution of the density profile as well, in the sense that the radial variation of the density profile shares the same general characteristics (discontinuities, asymptotics etc.) with the initial density profile we started with.

The stress-energy tensor is simply:

\begin{equation}
T_{\mu\nu}=(\rho+p)u_{\mu}u_{\nu}+pg_{\mu\nu}
\end{equation}
where $\rho$ is the total mass density of any matter constituents present and p is the pressure. Because of dark energy, the pressure is non-zero, and equal to $-\rho_{\Lambda}=-\frac{\Lambda c^2}{8\pi G}$. By manipulating the Einstein equations we obtain the following sufficient set of equations:
\begin{subequations}
\begin{align}
&\frac{\dot{A}^2}{A^2}+\frac{k(r)}{A^2}=\frac{G(r,t)}{A^3}\\
&2\frac{\ddot{A}}{A}+\frac{\dot{A}^2}{A^2}+\frac{k(r)}{A^2}=-8\pi p\\
&\frac{\partial p}{\partial r}=0\\
&\frac{\partial\rho}{\partial t}+\Big(\frac{\dot{X}}{X}+\frac{\dot{A}}{A}\Big)(\rho+p)=0\\
&X=\frac{\partial A/\partial r}{\sqrt{1-k(r)}}
\end{align}
\end{subequations}
where $k(r)$ is a suggestively written constant of integration to be determined by the initial conditions and the function G obeys the equation:
\begin{equation}
dG=-8\pi p(t)A^2 \frac{\partial A}{\partial t}dt+8\pi\rho(r,t)A^2\frac{\partial A}{\partial r}
\end{equation}
One sees already that this model predicts zero pressure gradients everywhere within the cosmological fluid, which sounds unphysical for an inhomogeneous metric like this. Focusing on the pressureless case we can reduce this
to only one equation that determines the evolution of $A(r,t)$:

\begin{equation}
\frac{\dot{A}^2}{A^2}+\frac{k(r)}{A^2}=\frac{F(r)}{A^3}+\frac{8\pi\rho_{\Lambda}}{3}
\end{equation}
where, if an initial condition of a uniformly expanding universe is assumed, F(r) obeys:
\begin{equation}
\frac{dF}{dr}=8\pi\rho_{M}A^2\frac{\partial A}{\partial r}, \hspace{1cm} \rho_{M}(r,t)\equiv \rho_{r,t}-\rho_{\Lambda}
\end{equation}
The latter equation justifies our choice to evolve every shell of the original matter distribution as an independent Friedmann-type universe, due to the fact that the function A(r,t) is not dynamical in the "radial" variable. For the sake of clarity, we will define a dimensionless function $a(r,t) = A(r,t)/r$ which tends to the scale factor $a(t)$ of the FLRW model.

The interesting question that arises here is how one determines the evolution of the mass density with time and how that compares with our model. Assuming a uniform rate of expansion at the time t=0, this evolution is simply found, given that the function F in equation
(B6) only possesses radial dependence:
\begin{equation}
\rho_{M}(r,t)=\frac{\rho_{M}(r,0)}{a^2(r\frac{\partial a}{\partial r}+a)}
\end{equation}
This shows that our cold dark matter approximation in appendix A is correct in the limit where the scale factor is uniform (more rigorously $d(ln a)/d(ln r)\ll 1)$. This
assumption holds to a very good approximation in the outskirts of the halo of a structure, and is therefore well justified.

Applying this machinery to a localized uniform spherically shaped overdensity, of density $\rho_1$ while the ambient density is lower and equal to $\rho_2$ we find that the description of the growing and collapsing structure is slightly different than the one we gave in appendix A with the corresponding curvature parameter reading:

\begin{equation}
\kappa(r)\equiv \frac{k(r)F(r)}{r^2}=\frac{1}{r^3}+\Big(\rho_{\Lambda}-\frac{3H_{0}^2}{8\pi G}\Big)\frac{1}{(\rho_1-\rho_2)r_{i}^3+\rho_2 r^3}
\end{equation}
However, it is easy to see that this equation becomes identical to the corresponding one derived in appendix A away from the structure core.


\begin{thebibliography}{}


\bibitem{ecf} R.V. Eke, S. Cole, and C.S. Frenk,
 Mon. Not. R. Astron. Soc. {\bf 282}, 263 (1996).

\bibitem{pfd} Pavlidou \& Fields 2005, DD paper

\bibitem{peeb84} P.J.E.Peebles, \apj {\bf 284}, 439 (1984).

\bibitem{peeb} Peebles, P.J.E. 1980, The Large Scale Structure of the
  Universe, Princeton Univ. Press, Princeton, NJ

\bibitem{krav} B. Diemer \& A.V. Kravtsov, \apj {\bf 789},1 (2014).

\bibitem{cuesta}  A. J. Cuesta, F. Prada, A. Klypin \& M. Moles, Mon. Not. R. Astron. Soc. {\bf 389}, Issue 1, pp. 385-397 (1996).

\bibitem{bertschinger} Bertschinger, E. 1985b, ApJS, 58, 39

\bibitem{gunngott} Gunn, J. E., \& J. R. Gott, 1972, \apj {\bf 176}, 1.

\bibitem{press} Press, W.H., Schechter, P., 1974, \apj, {\bf 187}, 425

\bibitem{white} White, S.D.M., Rees, M., 1978, Mon. Not. R. Astron. Soc., {\bf 183}, 341

\bibitem{nfw} Navarro, J.F., Frenk, C.S., White, S.D.M., 1996, ApJ, {\bf 462}, 563


\bibitem{halo} Tanvir, N. R. et al. 2012, MNRAS, 422, 162

\bibitem{sim} Lukic Z., 2008, Nonlinear growth of structure in cosmological simulations, Proquest Dissertations And Theses.

\bibitem{ante} Prada F., Klypin A. A., Simonneau E.,Betancort-Rijo J. ,Patiri S.,Gottlöber S., \& Sanchez-Conde M. A., 2005, \apj,{\bf 645}, 2

\bibitem{pavlidou} Pavlidou V., Tomaras T., Where the world stands still:
turnaround as a strong test of CDM cosmology, Journal
of Cosmology and Astroparticle Physics, Volume 2014,
September 2014 

\bibitem{krats}Surhud More, Benedikt Diemer, Andrey Kravtsov, The
splashback radius as a physical halo boundary and the
growth of halo mass, The Astrophysical Journal, Volume
810, Number 1 (2015)

\bibitem{splashback}O. N. Snaith, J. Bailin, A. Knebe, G. Stinson, J. Wad-
sley, H. Couchman, Haloes at the ragged edge: The im-
portance of the splashback radius, Monthly Notices of
the Royal Astronomical Society, Volume 472, Issue 3, 11
December 2017
\end{thebibliography}
\end{document}